\title{Determining the area of convergence in Bloodstain Pattern Analysis: a probabilistic approach}
\author{
        Francesco Camana, Ph.D.\footnote{Senior Technical Officer, Italian State Police (\emph{Polizia di Stato}), Interregional forensic science office (\emph{G.I.P.S.}), Piazzetta Palatucci, 5 - 35123 Padova (ITALY)}
}
\date{\today}

\documentclass[12pt]{article}
\usepackage{graphicx}
\usepackage{subfigure}

\begin{document}
\maketitle

\begin{abstract}
A complete procedure for identifying the area of convergence of blood drops originated from a single static source is presented. Both for bloodstains lying on an horizontal and on a vertical plane a complete study is developed, based on error analysis and on an opportunely defined joint probability density for the orientation of the horizontal projections of the trajectories of the drops. The method generates a probabilistic map for the area of convergence, directly linking the angles of impact, and their uncertainties, to the projection on the ground of the point of origin. One of the objectives consists in providing a statistical definition of area of convergence, extending to this topic the mathematical accuracy of the calculation of the angle of impact in bloodstain pattern analysis (BPA).
\end{abstract}

\begin{scriptsize}
\end{scriptsize}

\section{Introduction}\label{Introduction}
One of the most important results of the analysis of the bloodstains at a crime scene concerns the possibility of locating the area of origin of the projected blood. The drop colliding on an impact plane generates a bloodstain that is typically elliptical in shape. From the ratio of the axes of this ellipse the angle of impact $\alpha$ can be estimated by mean of the well established formula $\alpha=\arcsin {{d/a}}$, where $d$ and $a$ are the best estimations for the minor and major axis of the ellipse, respectively (see, \emph{e.g.},~\cite{JE, BG, BMR} and the references therein, with a particular note for the pioneer works of Balthazard \cite{BPD} and MacDonell \cite{MD1}). The angle $\alpha$ refers to the angle between the plane of impact and the tangent to the trajectory at the point of impact. The collection of the tangent lines at the point of impact, for a certain group of drops, is commonly projected on the horizontal plane and, whenever the bloodstains are referred to the same production event, these projections will converge and cross in an area that overlaps the normal projection of the region of origin \cite{JE,BG}. The importance of this result, from the forensic point of view, cannot be overestimated.

Looking at the scene from top, the path followed by the flying drops develops in straight lines\footnote{We neglect the unusual case where wind or other \emph{lateral} forces act on the drops.}; as a direct consequence, the procedure for the identification of the area of convergence is essentially a geometrical matter and requires little, if any, recourse to physical concepts and calculations. It's just a matter of drawing lines and spotting intersections.

This simple scheme is complicated by two main problems. First of all a mathematically satisfying definition of area of convergence is missing: the idea of \emph{area containing the intersections generated by lines \dots}~\cite{FBI} is not as precise as wished in this geometric context. The selection of the shape and of the dimensions of the area of convergence remains quite arbitrary and, apparently, it is not based on strict mathematical or geometrical considerations.

The second point concerns the uncertainties on the measurements of the bloodstains, originating a propagated error in the definition of the tangents of impact~\cite{WPDBR}. As a consequence of the analysis of the statistical errors, the intersections of the projections of the tangent lines at the point of impact will not result in single intersection points~\cite{R, BHC}, but more precisely in \emph{areas of probability of intersection}. How should this be taken into account?

The aim of this paper is to present a procedure that defines and identifies the area of convergence by mean of a statistical extension of the usual treatment of the uncertainties of the angles of impact of the bloodstains. 
The idea is to profit of the statistical errors in the estimation of the angles of impact to generate a planar \emph{probability map} for the intersections. The areas of highest probability of intersection will be interpreted as areas of highest probability for the origin of the projection. Every area in the plane will be therefore uniquely associated to a number: the probability that the area contains the normal projection of the origin in the horizontal plane.
In this way, the mathematical rigor that led to the estimation of the angles of impact will be transferred into the further steps of the analysis, within a coherent framework. The results of this procedure may also be the basis for the subsequent physical calculations of the trajectories of the drops and orient the search for the point of origin.

To achieve the goal, we will first recall in Sec.~\ref{angles_of_impact} the notions of tangent of impact, in the two different cases of stains deposited on horizontal or on vertical surfaces. In that section we will also show how to combine the traces found in these different planes into a unique context of analysis for the determination of the area of convergence.
The core argument will be presented in Sec.~\ref{area_prob}: by mean of the definition of a suitable joint probability density for the orientation of the projection of the tangents of impact in the horizontal plane, a statistical treatment will be argued for calculating the probability that specific areas of the ground contain the normal projection of the point of origin. Practical applicative examples will be described in Sec.~\ref{practical} and discussed in Sec.~\ref{discussion}.

\section{Angles of impact and uncertainties}\label{angles_of_impact}
The identification of the area of convergence preliminarily requires the analysis of the angles of impact of every single blood drop and this analysis is quite different for stains lying on horizontal or on vertical surfaces.

\begin{figure}
\begin{center}
\subfigure[]{
\resizebox*{6.55cm}{!}{\includegraphics{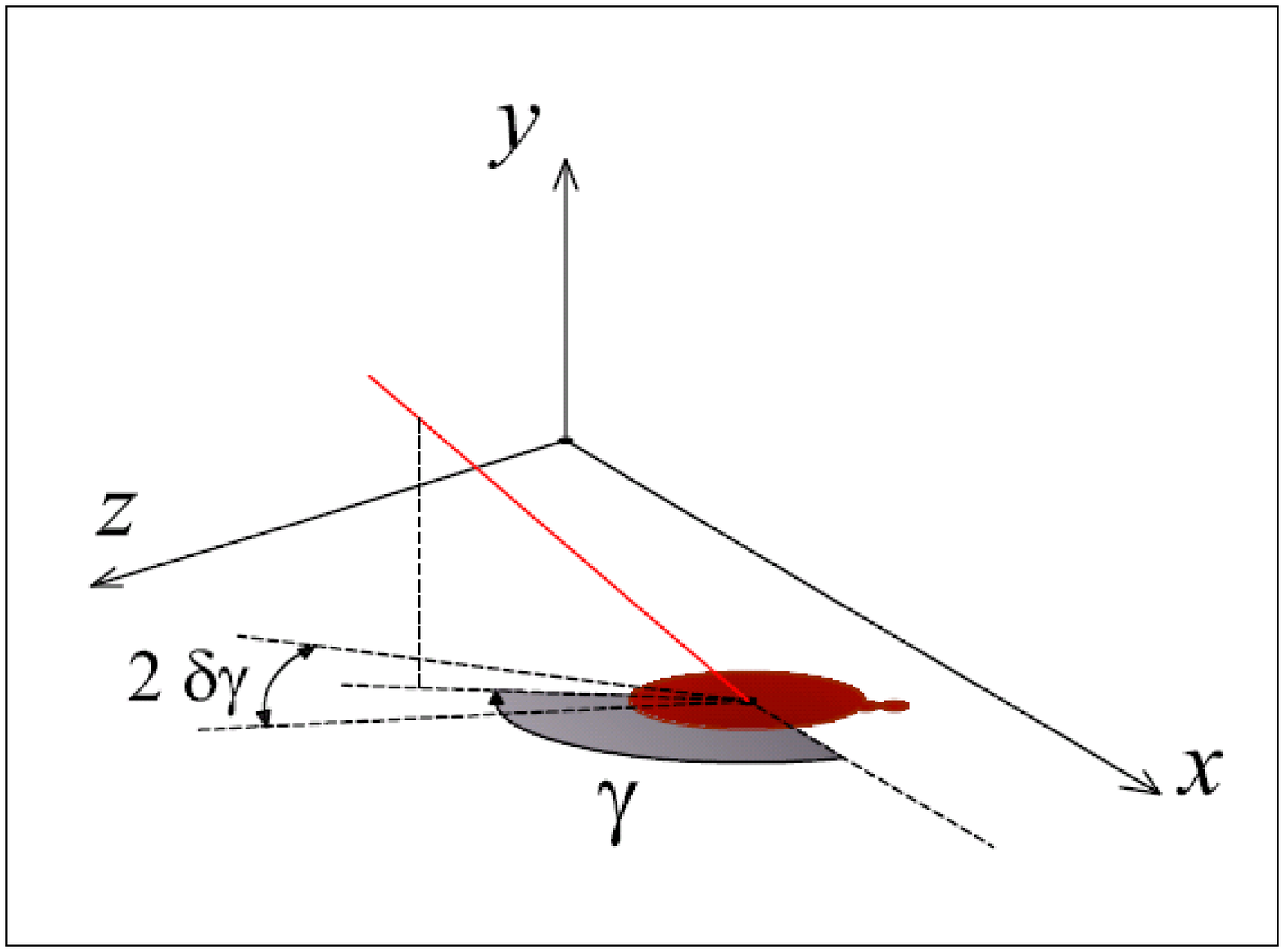}}\label{fig:gamma1}}%
\hspace{1mm}
\subfigure[]{
\resizebox*{6.55cm}{!}{\includegraphics{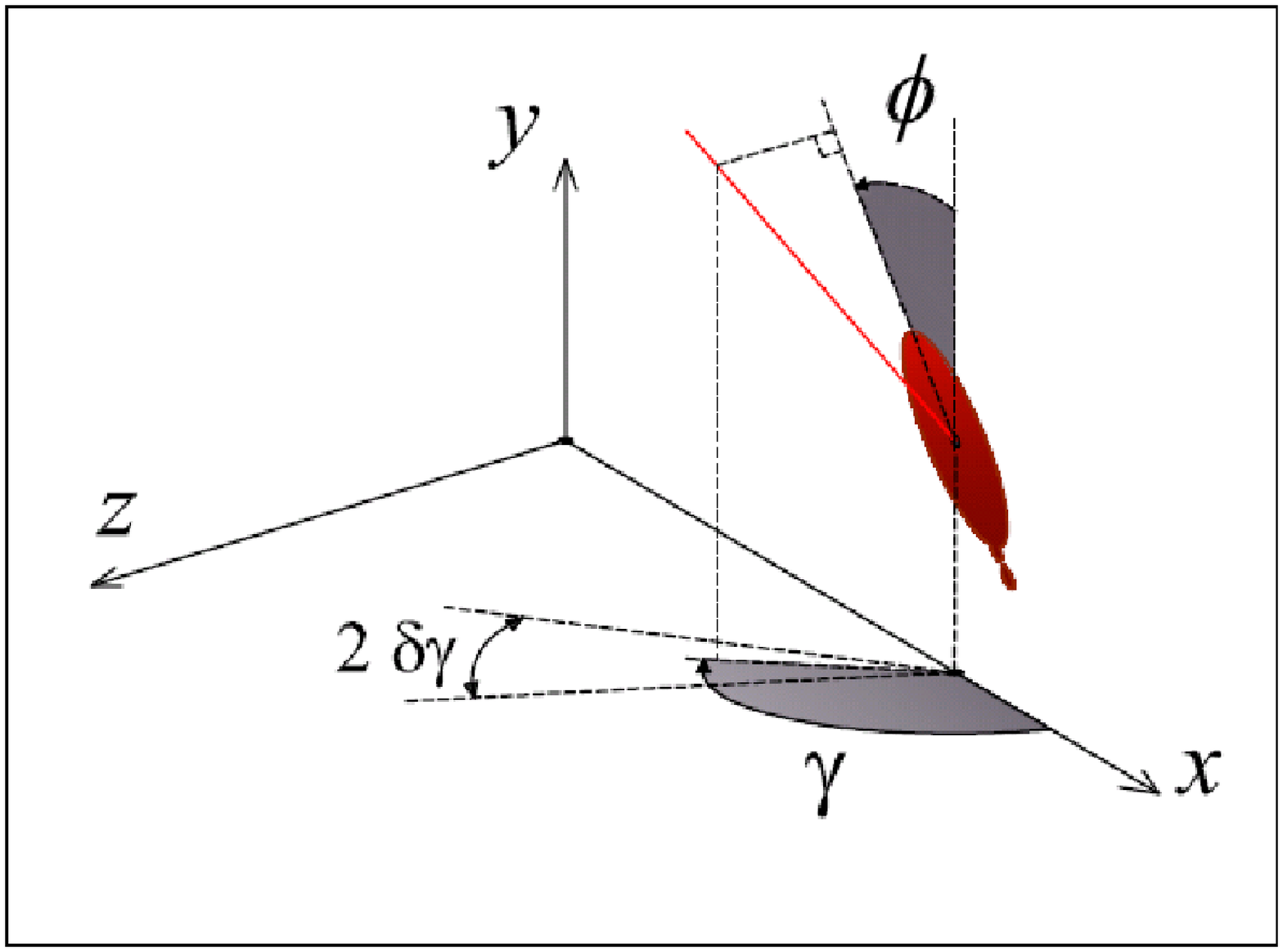}}\label{fig:gamma2}}%
\caption{The angle $\gamma$ between the $x$ axis and the normal projection in the horizontal plane of the tangent of the ballistic trajectory of the blood drop at the point of collision. In (a) the case of a stain on the horizontal plane is shown: the measurement of $\gamma$ is immediate. In (b) the stain lies on a vertical plane: the angle $\gamma$ must be calculated from the dimensions of the ellipse and the angle $\phi$ (see Eq.~\ref{eq_ang_gamma}). We decided to define $\phi\in{]-\pi,\pi]}$, negative as depicted in (b).}
\label{fig:gamma_angle}
\end{center}
\end{figure}

As a matter of fact, the elliptical bloodstains found on horizontal planes have the longest axis contained in the plane of the trajectory: in this case the identification of the plane of the trajectory is immediate, and simply demands the measurement of the angle of orientation of the stain, with respect to a certain reference. In this paper we will call this angle, with respect to the $x$ axis, $\gamma$ (see Fig.~\ref{fig:gamma1}). We decide to define $\gamma$ in the range $[0,2\pi[$. The uncertainty related to the process of measure of $\gamma$ is directly associated to an error $\delta\gamma$. In this simple case, $\delta\gamma$ is evaluated as standard deviation of a set of measurements of $\gamma$ itself or estimated from the precision of the measuring tool.

If the bloodstain is lying on a vertical surface, instead, some little work is required. The tangent of impact is not in general contained in the same plane as one of the axes of the ellipse (see Fig.~\ref{fig:gamma2}). To determine $\gamma$ (due to the presence of a vertical impact surface, $\gamma$ is here restricted in the range $]0,\pi[$) one must estimate the axes of the ellipse $d$ and $a$ and the angle $\phi\in{]-\pi,\pi]}$ between the vertical $y$ axis and the major axis of the ellipse (our choice is done so to have the angle $\phi$ negative in Fig.~\ref{fig:gamma2}). Then one can easily calculate the angle $\gamma$ by employing the trigonometric formula\footnote{Eq.~\ref{eq_ang_gamma} can be easily demonstrated by just applying the definition of the trigonometric functions, and considering that the sine of the angle of impact equals the ratio of the axes of the ellipse: $\sin\alpha=d/a$.}:
\begin{equation}
\gamma= \left\{ \begin{array}{ll}
\arctan {{{d\over a }}\over{ \sin \phi \sqrt{1-{{d^2}\over{a^2}}}}} & \textrm{if $a\neq d$ and $\phi\in{]0,\pi[}$}\\
\pi+\arctan {{{d\over a }}\over{ \sin \phi \sqrt{1-{{d^2}\over{a^2}}}}} & \textrm{if $a\neq d$ and $\phi\in{]-\pi,0[}$}\\
\pi /2 & \textrm{if $a=d$ or $\phi=0,\pi$.}
\end{array} \right.
\label{eq_ang_gamma}
\end{equation}
From Eq.~\ref{eq_ang_gamma} the error $\delta\gamma$ on $\gamma$ can be obtained. Following and extending the approach to the treatment of statistical error presented in \cite{WPDBR}, referring to the ISO 9000 standard \cite{ISO}, we may calculate the error $\delta \gamma$ as the square root of the variance of $\gamma$:
\begin{equation}
\delta \gamma=\sqrt { \left({{{\partial \gamma}\over{\partial a}}\delta a}\right)^2+\left({{{\partial \gamma}\over{\partial d}}\delta d}\right)^2+
\left({{{\partial \gamma}\over{\partial \phi}}\delta \phi}\right)^2},
\end{equation}
where $\delta \cdot$ is the uncertainty in the measure of ``~$\cdot$~" and where we have reasonably set the covariance to zero, between the three variables. The validity of this last hypothesis is evident after considering the independence of the three different measurements that are usually performed on a bloodstain, one angular and two orthogonal linear.

Passing through easy but cumbersome calculations, one finally obtains
\begin{eqnarray}
\delta \gamma=&&\sqrt {{{a^4\sin^2\phi}\over{\left(a^2-d^2\right){\left[\sin^2\phi \left(a^2-d^2\right)+d^2\right]}^2}}\cdot}\nonumber\\
&&\qquad\overline{\cdot\left[{{d^2}\over{a^2}}\left(\delta a\right)^2+\left(\delta d\right)^2+{d^2 \cos^2 \phi \over {a^4\sin^2 \phi}} {\left(a^2-d^2\right)}^2
\left(\delta \phi \right)^2\right]},
\label{eq_err_gamma}
\end{eqnarray}
for every $|\phi|\in {]0,\pi[}$ and $a\neq d$.
For $\phi=\pi/2$ one can easily check that Eq.~\ref{eq_err_gamma} reduces to Eq.~(20) of \cite{WPDBR}, in the case of zero covariance\footnote{Expressed with the notations of this article, the cited formula reads $$ \delta\gamma^2={1 \over{1-\left(d/a\right)^2}}\left({d\over a}\right)^2\left[{\left(\delta d\right)^2 \over d^2}+\left({d\over a}\right)^2{\left(\delta a\right)^2\over{d^2}}\right] ,$$ where the last, covariance term has been dropped, for the sake of clarity.}. Moreover, if $\phi \rightarrow 0$ then $\delta\gamma\rightarrow \delta\phi\sqrt{(a^2-d^2)/d^2}$, and the dependence on the values of $\delta a$ and $\delta d$ is suppressed, as expected. The rapid increment of $\delta\gamma$ for angles of impact around $\pi/2$, \emph{i.e.} for $a \simeq d$, is again an expected result, as in \cite{WPDBR}.

Despite its complexity, Eq.~\ref{eq_err_gamma} has a direct applicability and can be easily embedded in any software for the bloodstain pattern analysis\footnote{The software \emph{AnTraGoS}, realized by the author and employed by the forensic science service of the Italian \emph{Polizia di Stato}, already embeds the algorithm for the calculation of $\delta\gamma$ and of other angular uncertainties. \emph{AnTraGoS} has been developed and approved for internal use only.}.

Concluding this section, we point out that, from this point of the analysis on, the statistical uncertainties for the angle $\gamma$ can be treated in the same way, both for stains on the horizontal and on the vertical planes. The only difference rests on how these uncertainties have been measured or calculated.

\section{A probability density map for the area of convergence}\label{area_prob}
The next step of the analysis consists in putting together the bloodstains, with their angles of impact and relative uncertainties, in order to detect a possible area of convergence. Let $N$ be the number of drops (or, equivalently, stains) in analysis. After calculating the angles of impact, a collection of $N$ straight lines in the $xz$ plane is at hand: these lines originate from the stains, and each one is oriented at a $\gamma_k$ angle with respect to the $x$ axis (with $k=1, \dots, N$). These lines are the best estimations for the normal projection of the trajectory on the horizontal plane. For sake of simplicity, from now on, we will refer to these lines as $\gamma_k$-lines. Each $\gamma_k$-line is determined with an angular uncertainty given by $\delta{\gamma_k}$. According to the theory of random errors \cite{T} one can interpret the $\gamma_k$ angles and their uncertainties in a statistical sense. The idea is then to probe the points in the $xz$ plane in search of the areas of largest probability of intersection for the projections of the trajectories, given that their orientation can be associated to a certain statistical distribution.

\begin{figure}
\begin{center}
\subfigure[]{
\resizebox*{6.55cm}{!}{\includegraphics{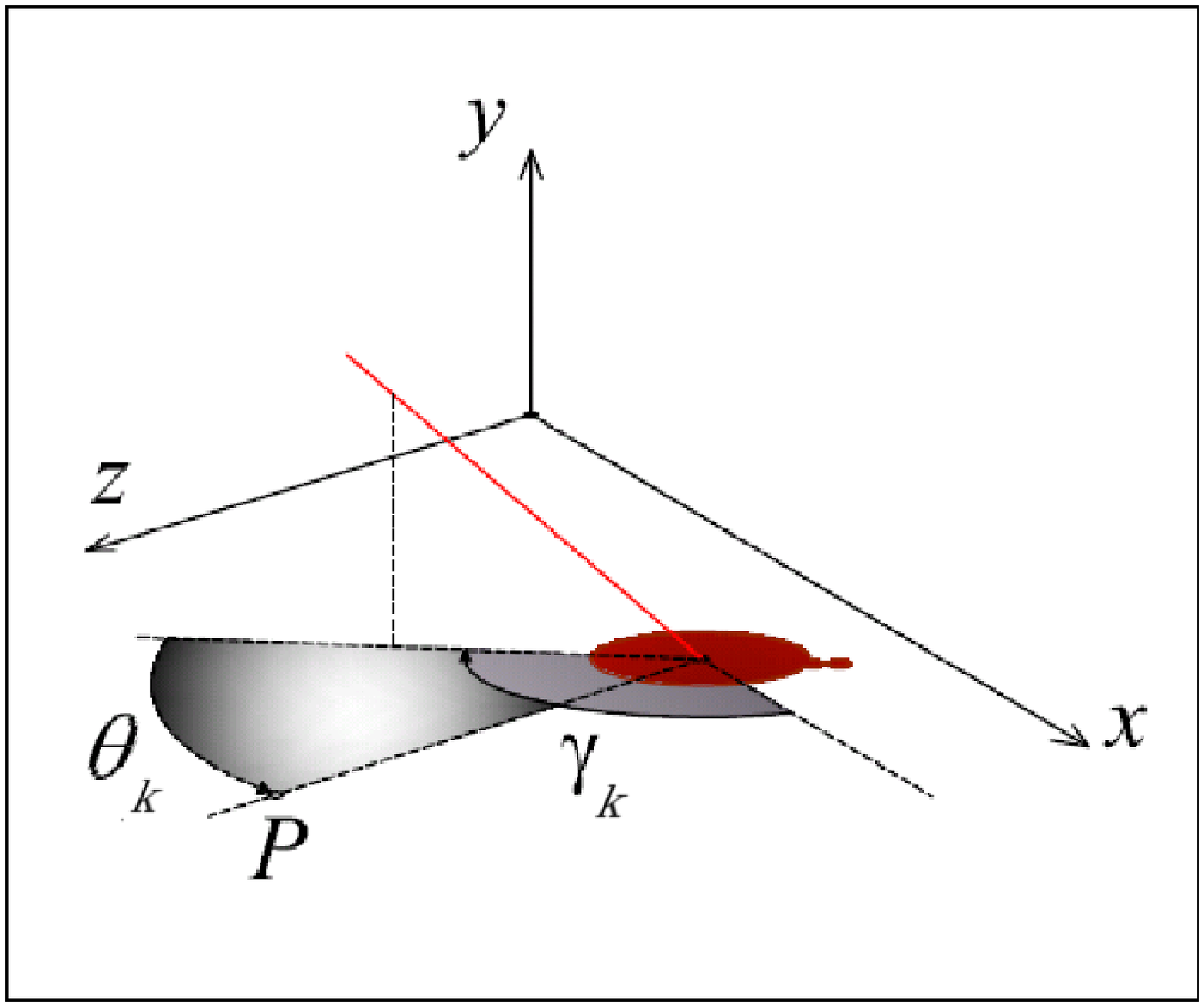}}\label{fig:theta1}}%
\hspace{1mm}
\subfigure[]{
\resizebox*{6.55cm}{!}{\includegraphics{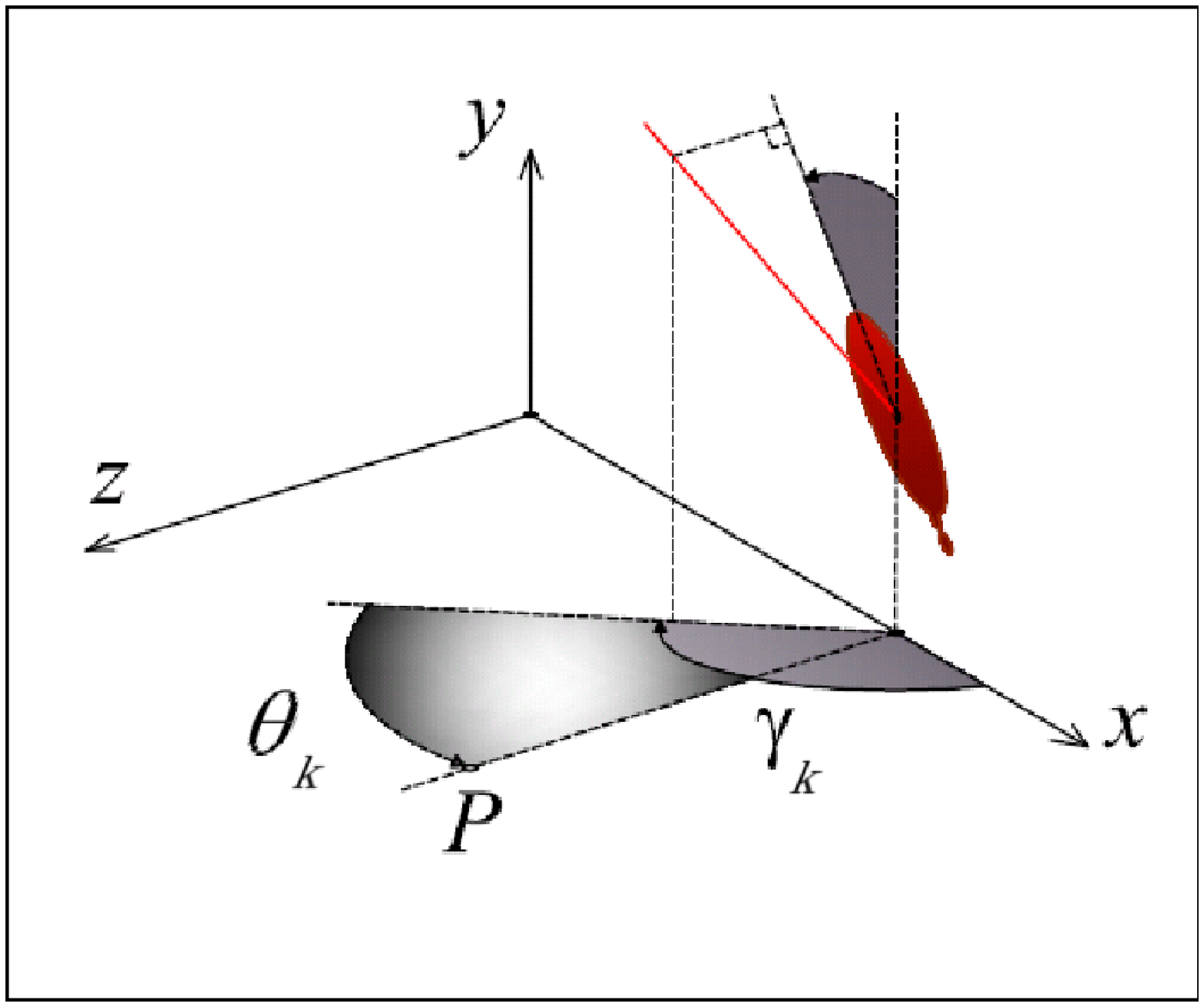}}\label{fig:theta2}}%
\caption{The angle $\theta_k$ between the $\gamma_k$-line and the line connecting a generic point $P$ of the $xz$ plane and the projection in the $xz$ plane of the point of impact. The only difference between the cases of deposits on horizontal (a) or on vertical (b) surfaces is that in the case (a) the projection of the point of impact and the point of impact itself coincide. In (a), for reasons of symmetry, the angle $\theta_k$ is selected in the range $]-\pi,\pi]$, and, if oriented as in the example figure, it is \emph{e.g.} chosen to be negative. In (b), due to the presence of a vertical surface, additional limitations on the values of $\theta_k$ must be imposed: in this case $\theta_k \in]-\gamma_k,\pi-\gamma_k[$.}
\label{fig:theta_angle}
\end{center}
\end{figure}

Referring to Fig.~\ref{fig:theta1}, we may imagine to pick a sample point $P$ on the $xz$ plane, which is supposed to be free of obstacles and obstructions, for simplicity. Connecting the point $P$ with the planar projection of the point of impact of drop $k$ a line is obtained, generating an angle $\theta_k$ with the $\gamma_k$-line ($\theta_k \in{]-\pi,\pi]}$).
Within the language of the stochastic processes applied to the error analysis \cite{VK}, we can assert that the probability density function (PDF) associated to the passage of the projection of the trajectory of the drop $k$ across the point $P$ is described by a wrapped normal distribution, the analog of the Gaussian in circular statistics \cite{F}. The form of this PDF is
\begin{equation}
\rho_k(P)=\rho_k(\theta_k)={{1}\over{\sigma_k \sqrt{2 \pi}}}\sum_{n=-\infty}^{\infty}\exp{\left(-{{{\left(\theta_k+2n\pi \right)}^2}\over{2\sigma_k}}\right)},
\label{eq_wrapped}
\end{equation}
with $\sigma_k^2\equiv{\left(\delta\gamma_k\right)}^2$ playing the role of a variance in the zero-mean distribution of the variable $\Theta_k$ \footnote{We denote here with capital letters the (stochastic) variables and with lower case letters the values assumed by these variables.}. In the usual case where $\delta\gamma_k\ll\pi$, it is sufficient to consider the central term of the sum (\emph{i.e.} the element with $n=0$) and the following approximation holds\footnote{The validity of this approximation is evident if one realizes that in this case the contribution to the wrapped normal distribution coming from the tails of the original, \emph{unwrapped} normal distribution is negligible. In the following, as an usual condition, we implicitly assume that $\delta\gamma_k\ll\pi$ and adopt a PDF in the form of Eq.~\ref{eq_PDF}.}
\begin{equation}
\rho_k(P)=\rho_k(\theta_k)\equiv \left\{ \begin{array}{ll}
{A_k\over {\sigma_k}}\cdot \exp \left(-{{\theta_k^2}\over{2{\sigma_k}^2}}\right) & \textrm{if $\theta_k\in{]-\pi,\pi]}$}\\
0 & \textrm{otherwise}
\end{array} \right.
\label{eq_PDF}
\end{equation}
with $A_k$ suitable normalization factor. 

These $k$ symmetrical distributions are approximately Gaussian, except for the cutoff imposed at $\pm \pi$ according to the limited range of the $\Theta_k$'s.
The case illustrated in Fig.~\ref{fig:theta2} can be treated in the same way but, due to the presence of an obstructing vertical surface of impact, the range of $\theta_k$ in Eq.~\ref{eq_PDF} is now restricted in the range $]-\gamma_k,\pi-\gamma_k[$. Other restrictions may be here imposed to take into account obstructions like furniture, walls, pillars, curtains, $\dots$

Now the idea is to join the PDF's of all the stains: for every test point $P$ of the $xz$ plane, one may evaluate the $\theta_1, \dots, \theta_N$ angles and calculate the joint probability density function $\rho_{joint}(P)\equiv\rho_{joint}(\theta_1, \dots, \theta_N)$ relative to the event that the point $P$ contains the horizontal projections of \emph{all} the $N$ stains. It is reasonable to assume the $\Theta_k$ variables as independent and therefore, by multiplying the $N$ PDF's, one finally gets
\begin{equation}
\rho_{joint}{(P)}=A\cdot\prod_{k=1}^{N}{1\over{\sigma_k}}\exp\left(-{{\theta_k^2}\over{2{\sigma_k}^2}}\right),
\label{eq_jPDF}
\end{equation}
where $A$ is an overall normalization factor determined by integrating $\rho_{joint}(P)$ over all the possible and compatible points of the horizontal surface in the scene. 

The probability ${Prob}~{(\Sigma)}$ that a given area $\Sigma$ contains the horizontal projection of the origin of the blood is then easily obtained by integrating this joint PDF over $\Sigma$:
\begin{equation}
{Prob}~{(\Sigma)}=\int_{P\in\Sigma}\rho_{joint}{(P)}{d\Sigma}.
\label{eq_Sigma}
\end{equation}

The integrals in Eq.~\ref{eq_Sigma} and for the determination of the normalization factor $A$ can be easily calculated by employing numerical methods.
In practice, the estimation of the integrals can be limited to areas around the points of intersection of the $\gamma_k$-lines, where the convergence is expected to occur. This last approximation is valid thanks to the rapid, exponential decay of the PDF's, far from the mean values.

\section{Practical use}\label{practical}
One of the most important advantages of the scheme presented in Sec.~\ref{area_prob} consists in the direct applicability of the method to an arbitrary large set of stains and in the easy realization of an automated procedure for the calculation process, by mean of suitable algorithms. In this section we present an example of application, realized on a simple forensic software.

First of all we notice that, even if the $\rho-$functions defined in Eq.~\ref{eq_PDF} and Eq.~\ref{eq_jPDF} are continuous in their range of validity, a discretization procedure is necessary for a computational approach. In the examples of this section, we have chosen, \emph{e.g.}, to sample the $\rho_k$'s and of the $\rho_{joint}$ functions at every $5~$cm, both in the $x$ and in the $z$ directions of the plane. The integrals are then calculated by supposing the $\rho$'s constant in the so defined $5~$cm$-$squares. The sampling and the integrations can surely be made more precise, but this level of approximation has proven to be sufficient in most real applicative situations\footnote{One may easily convince itself of this by considering that, on a typical circumstance, the distance between the area of convergence and the point of impact is of the order of 50~$-$~200 cm. At these distances the even low uncertainty of, $e.g.$, $3^{\circ}$ on the value of the angle of impact implies a linear range of uncertainty of $5-20$~cm for every single drop: at the end, when one considers the entire group of stains, this is also the order of magnitude of the uncertainty on the area of convergence. Moreover, spotting the area of convergence with a precision of 5~cm is more than sufficient to discriminate between mutually alternative situations that may have occurred in the scene; that is what really matters, from a forensic viewpoint.}.

\begin{figure}
\begin{center}
\subfigure[]{
\resizebox*{6.55cm}{!}{\includegraphics[trim=0 1cm 0 0, clip=true]{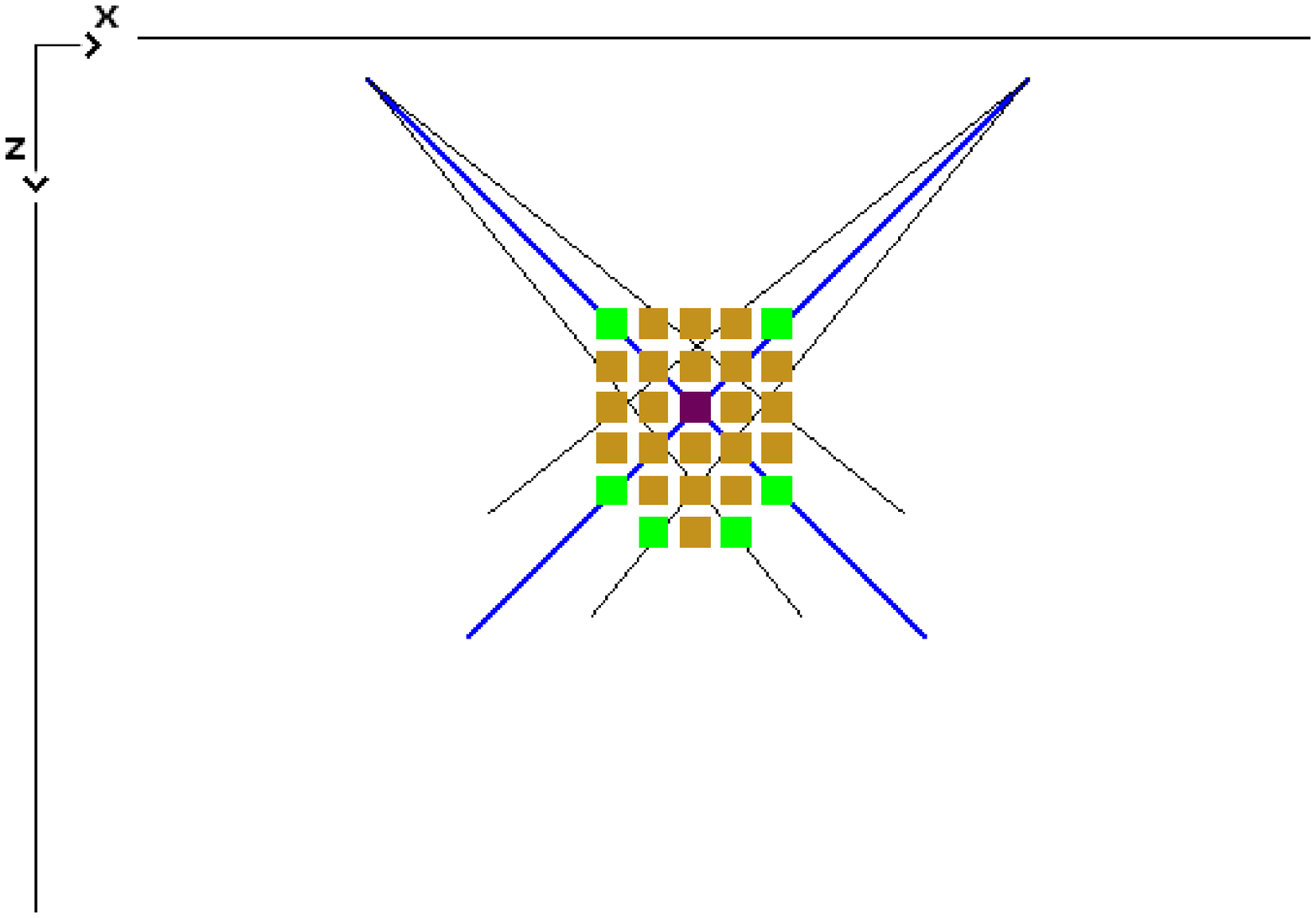}}\label{fig:acA}}%
\hspace{1mm}
\subfigure[]{
\resizebox*{6.55cm}{!}{\includegraphics[trim=0 1cm 0 0, clip=true]{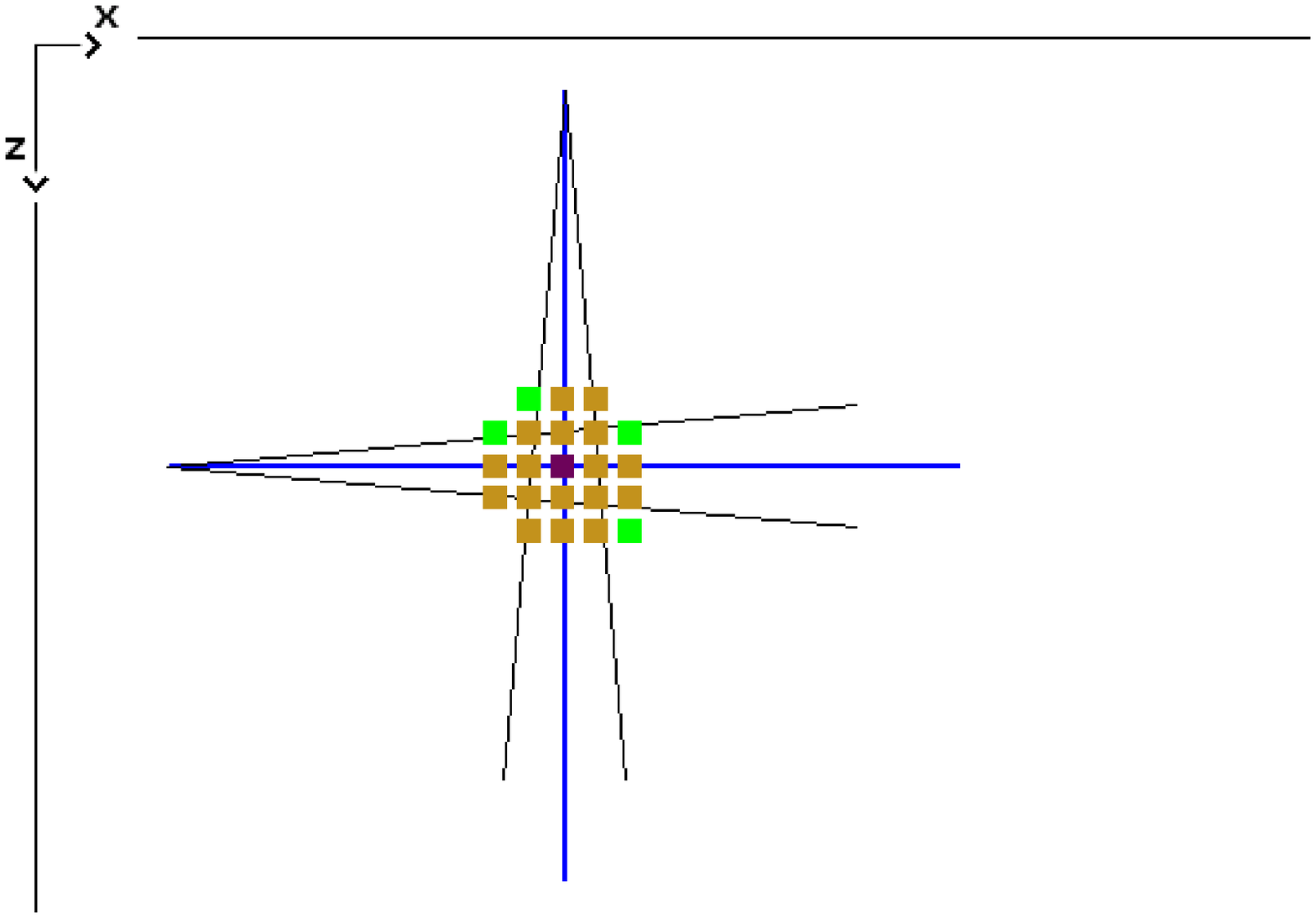}}\label{fig:acB}}%
\vspace{1mm}
\subfigure[]{
\resizebox*{6.55cm}{!}{\includegraphics[trim=0 1cm 0 0, clip=true]{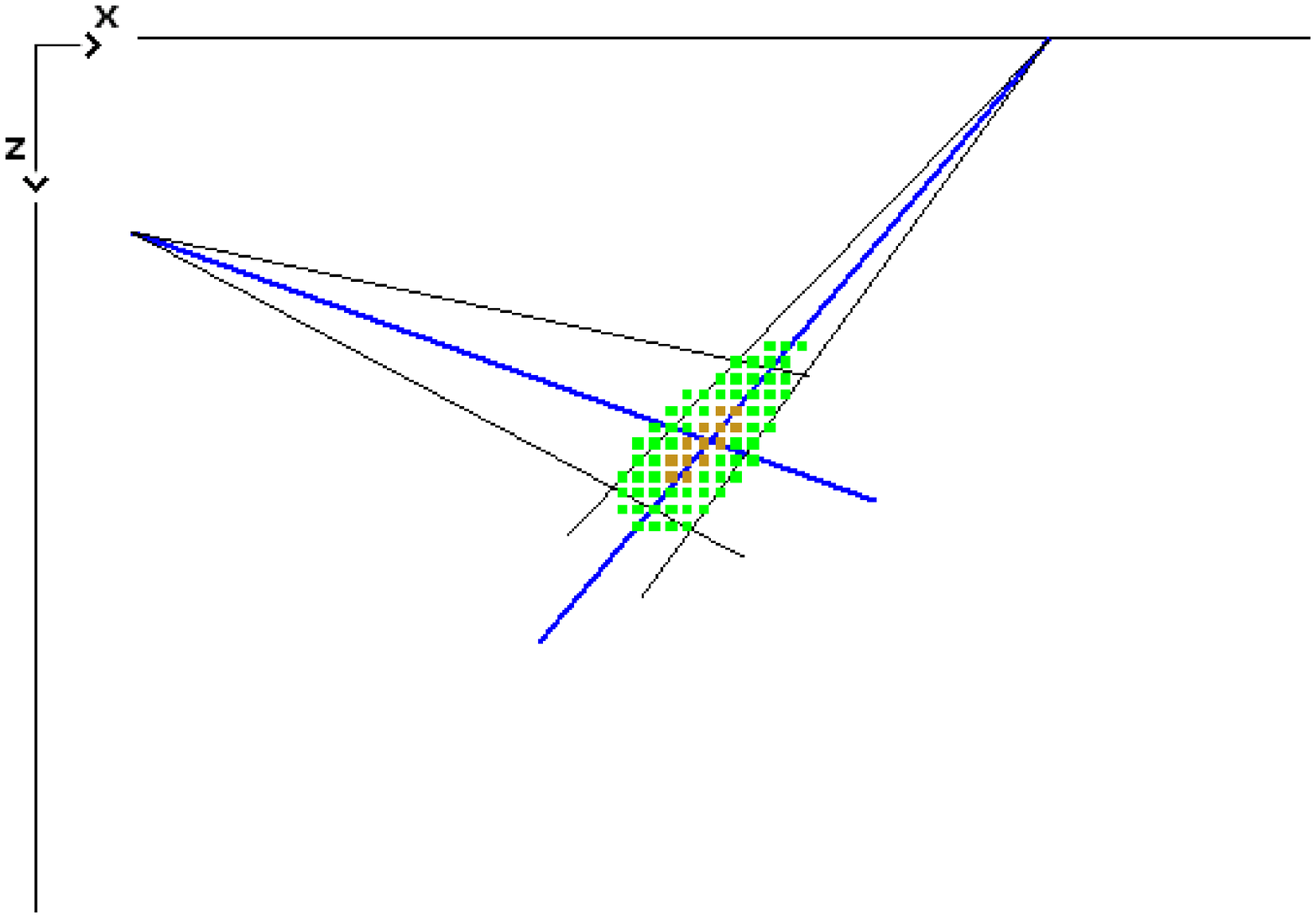}}\label{fig:acC}}%
\hspace{1mm}
\subfigure[]{
\resizebox*{6.55cm}{!}{\includegraphics[trim=0 1cm 0 0, clip=true]{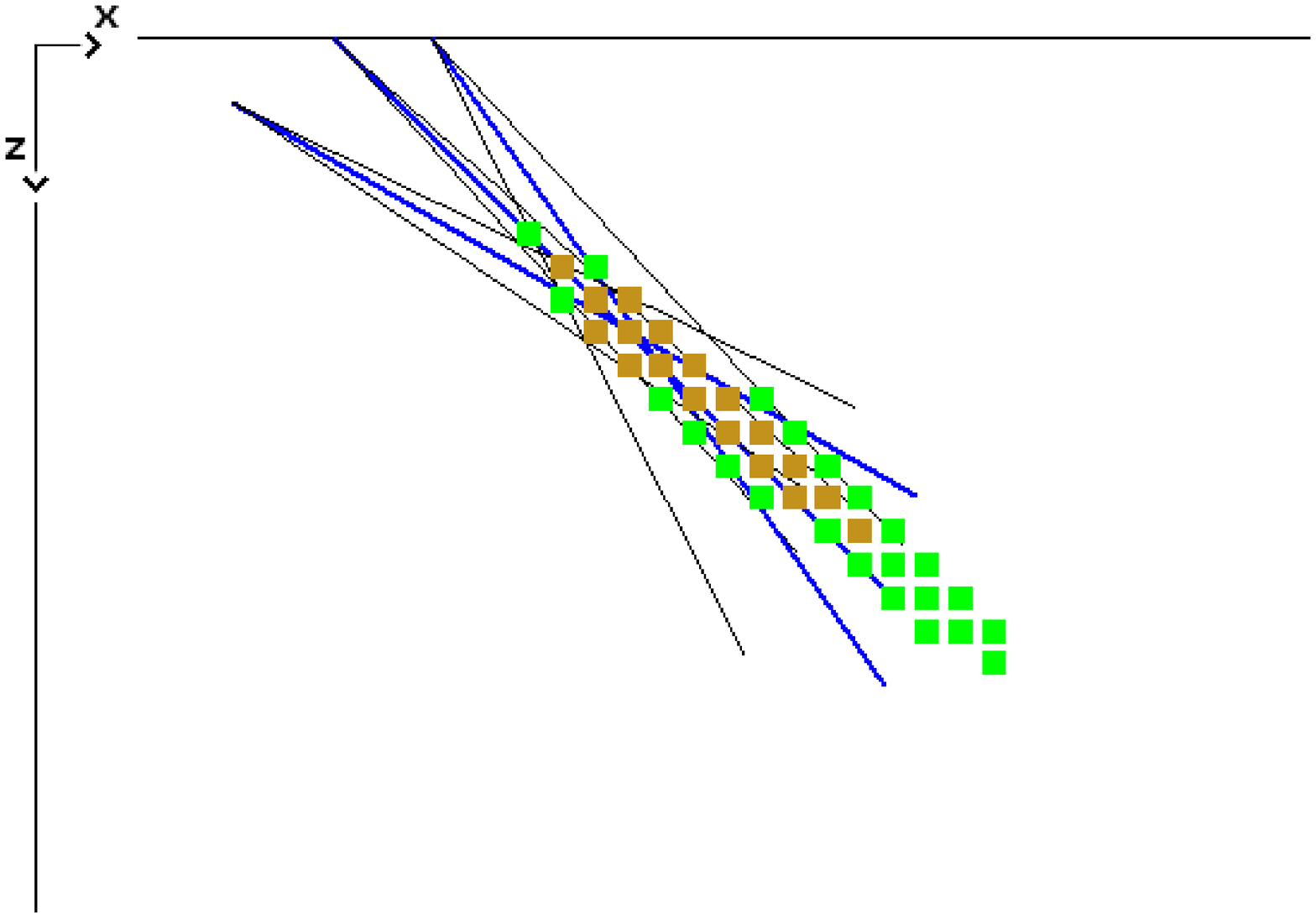}}\label{fig:acD}}%
\vspace{1mm}
\subfigure[]{
\resizebox*{6.55cm}{!}{\includegraphics[trim=0 1cm 0 0, clip=true]{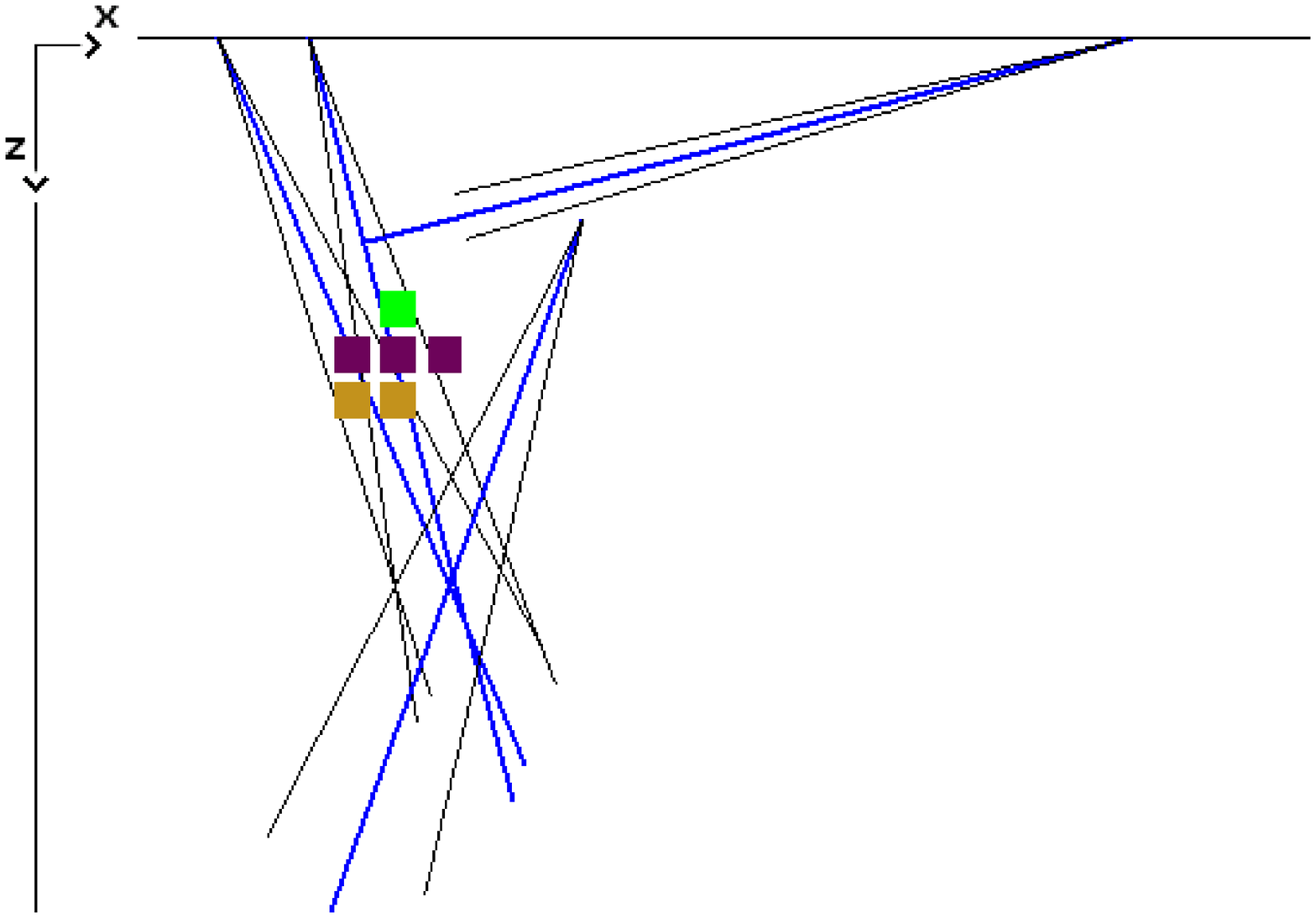}}\label{fig:acE}}%
\hspace{1mm}
\subfigure[]{
\resizebox*{6.55cm}{!}{\includegraphics[trim=0 1cm 0 0, clip=true]{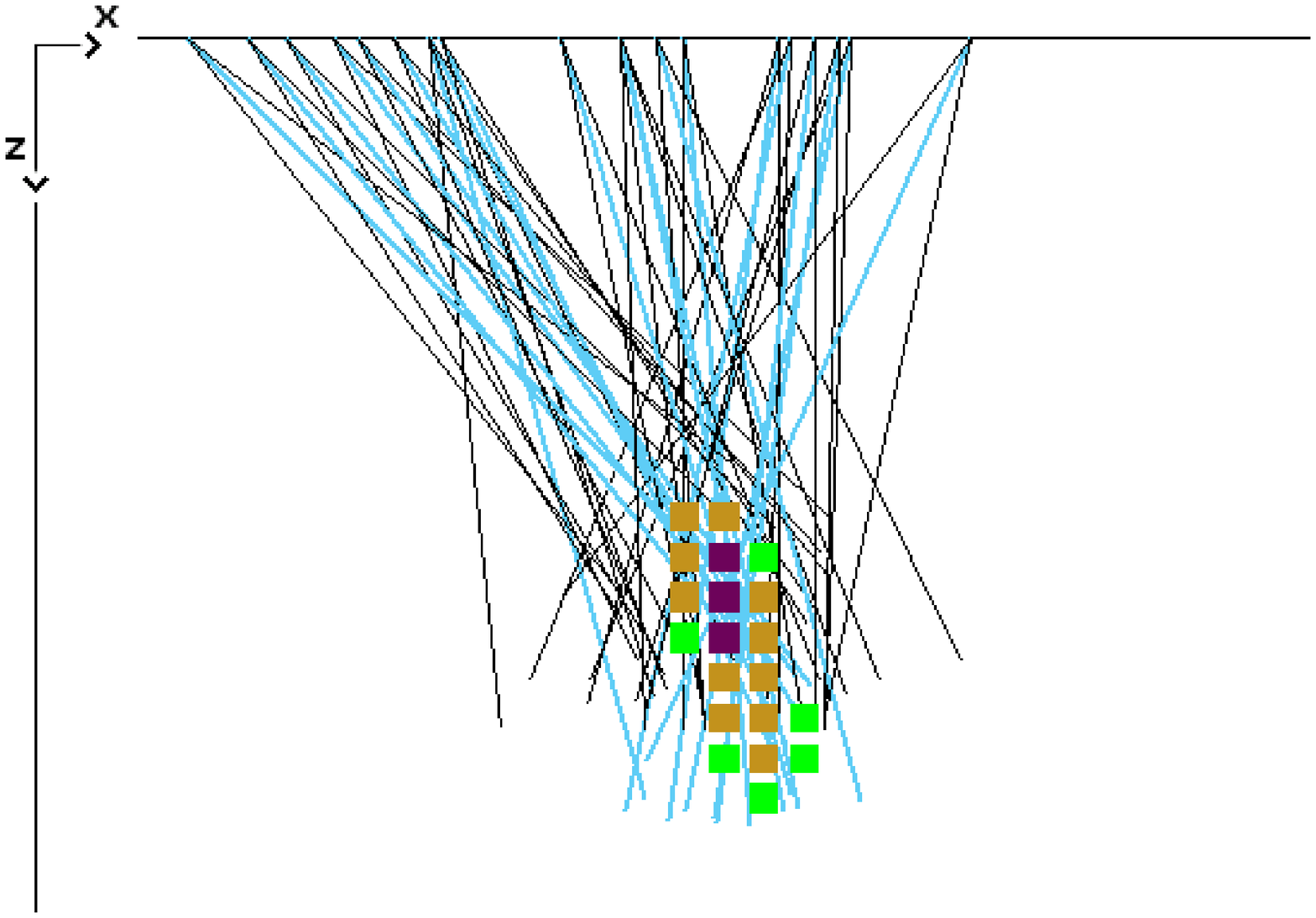}}\label{fig:acF}}%
\caption{Some applicative examples of the probabilistic method for the identification of the area of convergence, as implemented in \emph{AnTraGoS}, the software for the BPA of the forensic science office of the italian \emph{Polizia di Stato}. The color of the squares is related to the probability of joint convergence of the projected trajectories in the area: purple $>$10\%, orange 1--10\%, green 0.5--1\%. See the text for a detailed description.}
\label{fig:ac}
\end{center}
\end{figure}

Referring to Fig.~\ref{fig:ac} where different scales are used for convenience, the squares in purple cover areas where the probability of joint convergence exceeds 10\%, squares in orange are relative to probabilities in the range 1--10\% and squares in green to probabilities of 0.5--1\%.
The $\gamma_k$-lines are in blue and the uncertainty in their orientation ($\pm \sigma_k$) is in black. In the illustrative examples, the choice of the values of $\gamma_k$ and $\sigma_k$ is arbitrary.

The symmetric example in Fig.~\ref{fig:acA} refers to a simple case where the two stains have the same angular uncertainty. As expected, the calculated area of convergence is also symmetrical: the total probability associated to the colored area exceeds 95\%, while the probability of convergence in the purple square, where the $\gamma_k$-lines intersect, is only about 11\%. A slight non-symmetrical situation is instead depicted in Fig.~\ref{fig:acB}, where, due to the lower errors in the angles of impact, the resulting area of convergence is smaller (total probability $>$97\%, probability in the purple square 15\%).

In Fig.~\ref{fig:acC} the effect of an increased distance of convergence is shown. The $5~$cm$-$sampling is probably too fine at this scale and, even if the area of convergence is more regular in shape, it leaves out a lot of squares with probability less than 0.5\%: as a result, the total probability of the colored area is only about 61\%.

An important example is presented in Fig.~\ref{fig:acD}: here the three $\gamma_k$ angles have more or less the same value. Even if the $\gamma_k$-lines converge in a relatively small area, the presence of an error in their determination results in a very large area of probability of convergence, oriented along the direction of these lines. The colored area contains only 48\% of the total probability of convergence, indicating that a large error in the determination of the point of origin is to be expected. The example illustrates the problems related to the standard definition of area of convergence \cite{FBI} and also highlights the importance of having differently oriented stains in the scene, to have a reduced overall error in the analysis (see, \emph{e.g.}, Sec. 3 of \cite{R} and \cite{BSL}).

In any case, the validity of the assumptions of the method should be always carefully verified, as shown in the problematic case of Fig.~\ref{fig:acE}, where the algorithm has calculated an area of convergence despite of the evident presence of stains belonging to different production events. The fact that the spotted area of convergence lies far from the effective area of intersection of the $\gamma_k$-lines can be indicative on an error in the assumption of single projection. All the three angles for the stains to the left converge in a small area, while the stain to the right does supposedly belong to another projection: a clear indication that its study should be separated from the analysis of the other three. It is worth mentioning here that it is particularly easy to include/exclude single stains in the procedure and to check the effect of this inclusion/exclusion in the results: this recursive testing method can be of some help to identify the groups of stains with common origin.

The efficiency of the method and of its implementation are evident in the combined treatment of large numbers of drops (up to the order of $10^2$), as in the most common forensic cases. As soon as the angles are calculated with their errors, the computational time for the solution of the integral in Eq.~\ref{eq_Sigma} is practically zero and the area of convergence can be immediately visualized. An example taken from a real case is depicted in Fig.~\ref{fig:acF}: the purple area is representative of a probability of about 49\%, while the total probability of the colored area amounts to 99\%.

\section{Discussion}\label{discussion}
The described statistical approach can locate the areas of the horizontal plane where the highest levels of probability of convergence are concentrated. 
From these values of probability suitable, quantitative definitions of area of convergence may accordingly be stated. One may, \emph{e.g.}, choose a level of probability $\overline{p}$, close to unity, and define the area of convergence $\overline{\Sigma}$ as the smallest of the surfaces $\Sigma$ of the horizontal plane where the relation $Prob\left(\Sigma\right)\geq \overline{p}$ holds. 
Different choices for the value of $\overline{p}$ spot areas that obviously may differ in shape or size, and a certain variability can also be ascribed to the arbitrariness on the discretization of the horizontal plane for the numerical estimation of the integral in Eq.~\ref{eq_Sigma}. As noted above, the accuracy of a discretization with 5~cm spacing is usually more than acceptable, for all practical purposes.

This scheme, apart from being a natural extension of the error analysis in BPA, provides a solution to the important issues and problems connected to the standard definition of area of convergence, like the undetermined connection between areas and points of intersection. As seen in the comment to Fig.~\ref{fig:acA}, the small area where the $\gamma_k$-lines intersect is effectively related to a low probability of convergence, and a wider area has to be considered to get higher values: an area that approximately covers the uncertainties, too. 

The application of the procedure to specific situations like the one depicted in Fig.~\ref{fig:acD}, where, due to the similar orientation of the angles of impact, the intersections of the $\gamma_k$-lines are not indicative for the convergence, also reveals the importance of the probabilistic approach: the area of convergence is effectively quite widespread, in that case. As seen in this and in other examples, the area of convergence does not usually approximate a circular shape: this evidence is indicative of the necessity of a statistical approach that, like the one proposed, may account for the asymmetries of the specific situation.

The stains with lower errors are, as expected, the most relevant in the definition of the joint probability density of Eq.~\ref{eq_jPDF}: great care should then be used in minimizing the errors in the previous steps of the BPA, \emph{i.e.} in carefully selecting the stains, taking the photographs and estimating the best fitting ellipses. Not surprisingly, large variances $\delta\gamma_k$ ultimately result in the uselessness of the drops in the overall analysis, especially in presence of many other stains with significantly lower variance.

The idealized situation sketched in this article can be improved by setting suitable contour conditions that account for the geometrical limitations imposed, \emph{e.g.}, by the presence of furniture, walls or other obstructions: every single drop might have its own limit values and the overall consistency of the hypotheses should be always carefully checked. Particular attention should be paid in the implementation and in the use of algorithms that automatically generate a probability map for the area of convergence: the results might not be precise enough to describe a specific situation, and in most cases a \emph{manual} check of the single PDF's is suggested.

A possible extension consists in pushing the error analysis into the field of the three-dimensional reconstruction of the trajectory of the blood drop. Even if the idea seems more than reasonable at this point \cite{BKNAST, CH}, some complications arise from the necessary introduction of other variables such as the (quadratic) air friction and the impact velocity of the drop: these physical quantities are generally estimated at a crime scene with a relevant uncertainty \cite{GK,HSMC,KD,K}. The simple scheme described in this article can hardly have a similar counterpart in a three-dimensional context and the smart solution proposed in \cite{CH} is however limited to the hypothesis of linear drag, which is not properly the case of the blood drops originated in an usual crime scene.

\section{Conclusion}\label{conclusion}
In Sec.~\ref{angles_of_impact} we have presented a procedure for grouping the stains, deposited on horizontal and on vertical surfaces, into a single analytical framework. The most important result, derived from error analysis, is enclosed in Eq.~\ref{eq_err_gamma}, that provides an estimation of the statistical uncertainty in the determination of the horizontal projection of the tangent of impact. 

By extending the error analysis beyond the determination of the angle of impact, we could then construct in Sec.~\ref{area_prob} a simple applicative scheme for a statistical characterization of the area of convergence and for a quantitative definition of its shape and extension. The method is based on the definition of a joint PDF for the orientation of the horizontal projections of the tangents of impact (Eq.~\ref{eq_jPDF}) which forms the basis for the measurement of the probability of convergence in a given area (Eq.~\ref{eq_Sigma}). We have specified the possibility of inserting the procedure into a simple software to perform the described analysis in very short times.

We have also shown how the model works in different situations and the advantages of this statistical treatment.
As a side result, we could describe how some applicative examples support other previously published results in BPA: the errors in the determination of the area of convergence become larger when the drops are all in the same impact plane or when the projected trajectories intersect at a small angle.

Future work may concern the spotting of the point of origin and of its statistical uncertainty in 3D: the task appears as necessary as complex. 
The necessity to include non-linear effects in the dynamics generates problems for an analytical solution: for all practical purposes, a good estimation of the minimum and maximum for each linear dimension of the point of origin can now be obtained by repeatedly calculating the time-reverted equations of motion for each drop \cite{L, BHC}, selecting time by time different values for the key physical quantities. 
However, our belief is that an extended and coherent error analysis should be employed also in this last stage of the BPA. In this way the general framework would certainly be more consistent, as well as the forensic relevance of the results.

\section*{Acknowledgements}
I acknowledge Massimiliano Gori for his incentive and for the useful discussions. I also acknowledge Fulvio Baldovin for reviewing the text and for his precious comments on circular statistics. I finally acknowledge the Italian \emph{Polizia di Stato} for the granted permission.
\bibliographystyle{abbrv}


\end{document}